\title{Toward Applying Quantum Computing to Network Verification}

\documentclass[10pt,twocolumn]{article}
\usepackage[letterpaper, margin=0.75in]{geometry} 
\usepackage{amsmath, amssymb} 
\usepackage{graphicx} 
\usepackage{subcaption} 
\usepackage{authblk} 
\usepackage{fancybox} 
\usepackage{hyperref} 
\usepackage{balance} 
\usepackage[numbers]{natbib} 

\providecommand{\Description}[1]{}
\title{Technical Report: Toward Applying Quantum Computing to Network Verification}

\author[1]{Kahlil Dozier}
\author[1]{Justin Beltran}
\author[1]{Kylie Berg}
\author[1]{Hugo Matousek}
\author[1]{Loqman Salamatian}
\author[1]{Ethan Katz-Bassett}
\author[1]{Dan Rubenstein}
\affil[1]{Columbia University, New York, NY, USA}

\newcommand{\TROnly}[2]{#1}

\newcommand{\ket}[1]{\ensuremath{\left| #1 \right\rangle}}

\newenvironment{myitemize}
{\begin{list}{$\bullet$}{
    \setlength{\topsep}{5pt}
     \setlength{\labelwidth}{0em}
     \setlength{\labelsep}{1em}
     \setlength{\itemsep}{0pt}
     \setlength{\leftmargin}{0em}
     \setlength{\rightmargin}{0cm}
     \setlength{\itemindent}{1em} 
   }
  }
{\end{list}}

\begin{document}

\balance

\maketitle
\section*{Technical Report Note}
This document is an extended technical report based on our original paper, \emph{Toward Applying Quantum Computing to Network Verification}, published in the proceedings of the 23rd ACM Workshop on Hot Topics in Networks (HOTNETS '24)~\cite{original-paper}. The report provides additional technical details and supplementary analysis that complement the results presented in the published version.

\begin{abstract}
Network verification, broadly defined as proving the correctness of certain properties resulting from a network's configuration, cannot be efficiently solved on classical hardware via brute force.  Prior work has developed a variety of methods that scale by observing a structure in the search space and then evaluating classes induced by that structure.  However, even these classification mechanisms have their limitations.  In this paper, we consider a radically different approach: applying quantum computing to more efficiently solve network verification problems.  We provide an overview of how to map variants of verification problems into unstructured search problems that can be solved via quantum computing with quadratic speedup, making the approach feasible in theory to problems that twice as big in the size of the input.  Emerging quantum systems cannot yet tackle problems of practical interest, but rapid advances in hardware and algorithm development make now a great time to start thinking about their application.  With this in mind, we explore the limits of scale of the problem for which quantum computing can solve network verification problems as unstructured search.
\end{abstract}

\section{Introduction}
\label{sec:Intro}
Networks are designed with the intention of meeting certain goals and requirements.  Network verification refers to a class of techniques used to check that various properties of concern to network administrators are satisfied by a network's configuration. For instance, common concerns might be to ensure that routes between specific sources and destinations explicitly do (or do not) pass through specific intermediate routers, or that the number of hops is bounded below some threshold (precluding infinite loops).  There is a significant body of prior art that explores verification in the context of properties of different networks using different techniques to efficiently perform the verification process~\cite{kazemian2012header, netdice, minesweeper}. Despite their differences, the focus of each verification technique can generally be split into two classes:

\begin{myitemize}
    \item Data plane: whether the forwarding table rules and other data plane elements (e.g., ACLs)
    achieve desired properties for how data flows through the network. 
    \item Control plane: whether the configuration of routing protocols results in data planes that satisfy desired properties.
    
\end{myitemize}

Substantive progress has been toward the design of approaches that address many concerns of network administrators, despite the problems' inherent complexity~\cite{surveyPaper}. However, there remains a large set of properties for which existing approaches do not offer efficient solutions.  For instance, for data plane verification, it remains elusive to determine whether routing paths through the network are guaranteed to be bounded by a fixed number of hops or fixed delay.  For control plane, evaluating whether desired properties hold across flows in the face of link failures remains an open challenge.

Network verification differs from a traditional single-machine program verification in three important ways:

\begin{myitemize}
    \item The network ``program'' is implemented across a distributed set of components (e.g., routers).  While what each component does is relatively simple when compared to a standard (centralized) program, the complexity in analyzing properties of a network system greatly depends on how these relatively simple parts interact as a whole.
    \item The network ``program'' is generally an operation that needs to be completed quickly (e.g., transit from a source to a point), and loops or significant recursion is highly undesirable.  In contrast, traditional programs often rely heavily on long loops and recursive calls to effectively implement operations of significant complexity.
    \item When a property is violated, the verification system should (hopefully) return an {\em input} that caused the violation.  Inputs in traditional programs are inputs to the program and are often themselves quite large, for example files or large arrays.
    In contrast, network verification inputs for a given network tend to be small, on the order of tens of bits: e.g., a packet headers or a set of failed links that cause a property to be violated.

\end{myitemize}

Because in network verification the internal state of the system may be highly complex but an input of the problem is small, quantum computing may offer a more efficient means at solving a variety of problems that remain elusive classically.  While quantum computing technology is not yet at a point where it can be deployed to address reasonably-sized networks, the trajectory of advances in the field is moving so quickly that now is a good time to further our understanding of how to apply quantum technology so that we can rapidly build solutions when its applicability does reach fruition.

{\bf Classical ``structured'' v. Quantum ``unstructured'':} For classical solutions to be efficient, they generally require being able to make assumptions about a given network, effectively applying ``structure'' to inputs.  Such a structure allows inputs to be grouped into (evolving) classes, where the classical approach analyzes the inputs on networks by each class, one at a time.  Classical techniques become intractable when the inputs lack this structure necessary to separate them into a small number of classes.  In contrast, quantum computing is known (in theory) to offer quadratic improvement with respect to classical computing when dealing with ``unstructured'' data: this allows a doubling of the size (number of bits) of the input space, where doubling can be beneficial in practice (e.g., handling headers twice the size).

In this paper, we show how to map variants of previously considered network verification problems into a quantum computing framework.  The variants we consider are difficult to solve efficiently on classical infrastructure because of their inherent complexity \cite{NVComplexity} but can be solved with quadratic speedup in a quantum computing context as unstructured search problems by applying Grover's Algorithm \cite{grover1996fast}. These results are merely proof-of-concept, as hardware is not ready to implement these algorithms.  Also, quantum computing may in fact offer significantly more computational power than what we demonstrate here \emph{if} one can find some structure in the data whereby a solution beyond Grover's Algorithm may be applied with significantly greater speedup.  We demonstrate one minor enhancement technique for our control plane example known as {\em amplified Grover} that can bias solutions toward those with a desired number of lossy links~\cite{BrassardHoyer}.

In the remainder of this paper, we present a formulation of the general network verification problem within the framework of an unstructured search scenario and illustrate how to use Grover's algorithm to address some of the challenges. Because it is hard to exactly assess practical costs (error correction and the stochastic nature of quantum measurement), it is difficult to precisely assess how long our proposed unstructured searches will take.  For now, we limit our scalability analysis to the number of qubits (the analog of classical bits) needed to perform verification as unstructured search on emerging quantum devices.





\section{NWV Formalism}
\label{sec:NWVFormalism}

In this section, we introduce the basic nomenclature we use to define a NWV problem, which is effectively a 4-tuple consisting of 1) the network components, 2) the protocols run on those network components, 3) instances of interest, and 4) properties to verify.  A NWV system incorporates these four components by emulating the protocols on a software manifestation of the network and determining,  across the set of instances of interest, which instances satisfy the properties to verify and which do not:

\begin{myitemize}
\item {\em Components} are the network hardware, i.e., routers, switch\-es, links, that must be emulated in the NWV system.

\item {\em Protocols} are the software run on the components.  Much of the innovation in NWV involves translating from the underlying protocol that would function at any time on an individual instance to a form that functions on the (structured) classes of inputs.  For instance, in the data plane, a protocol run at a router may determine a packet's next hop router based on a packet header.  In a NWV system \cite{kazemian2012header}, a router would process a data structure representing a set of packet headers and indicate for each next hop router the subset of these headers that would transition there.  In the control plane, a router's choice of a next hop for destination-based routing will be dependent on the set of network links that have failed.  A NWV system \cite{netdice} would process a data structure that tracks the likelihood of a specific next hop being taken as a function of distribution on which network links are assumed to fail.

\item {\em Instances} of interest describe the scenarios over which NWV is performed, e.g., in the data plane, the set of possible packet headers, and in the control plane, the set of possible link failure combinations.

\item The {\em property} is the specific condition that the verifier is to identify instances that either succeed or fail to satisfy the property.  In the data plane, the property might be to find the sets of headers that never reach an intended destination (infinite loop) or whose path is beyond a reasonable length.  In the control plane, it may be a path either proceeding or failing to proceed through some intermediate router.
\end{myitemize}

In classical approaches, there are simply too many instances (e.g., possible packet headers, combinations of network links that fail) to test for.  Hence, prior art has found a variety of ways to find "structure" within the problem such that rather than analyze individual instances, instances can be classified into what is hopefully a much smaller number of classes.  Classes can be dynamically modified during the evaluation process, but as long as the number of classes remains small, the problem can be solved in a practically reasonable time.  For instance, in \cite{kazemian2012header}, sets of headers that are equivalent with respect to the checked property can be grouped together as a union, intersection, or difference of regular expressions of their bit patterns.  As these classes move between routers and headers are modified, these modifications are captured by recomputing the corresponding resulting combinations of regular expressions that capture the change.  In \cite{netdice}, for a given source-destination pair, links can a priori be identified as "cold" and can be omitted from the evaluation when it can be proved that the likelihood of enough links failing such that the cold link would be part of the routing path from that source to the destination is inconsequentially small.

\section{Quantum Unstructured Search Background}
\label{sec:QuUSearch}
In this section, we present a very high-overview of how {\em unstructured search} is implemented in a quantum computing context using Grover's algorithm, which has been shown to have quadratic speedup over classical analogs \cite{grover1998framework}.  
Our explanation provides a minimal description of Grover's algorithm sufficient to support our discussion in \S\ref{sec:MappingNVtoQuantum}.

{\bf Grover's algorithm} is most easily described in the context of a problem involving a {\em black box function} $f$ whose internals are presumed unknown.  The function $f$ itself takes an $n$-bit input and based on the input, deterministically outputs either a 0 or 1, where an output of 1 is uncommon.  $f$ itself is assumed to have no structure, i.e., knowing the outcome of a subset of the inputs provides no information about which as-of-yet evaluated inputs might yield a 1.  The goal is to find an $n$-bit input whose output evaluates to 1 (if such an output exists).  

In a practical use, the function $f$ in Grover is not an unknown oracle but is a reasonably computable (i.e., polynomial-time) {\em verifier} whose use is similar to that of a verifier in a nondeterministic Turing machine.
It is assumed that building the verifier function is easy.  However, to find a solution, the verifier must be run on a large (often exponential) number of inputs.  Classical approaches require $O(2^n/k)$ time when $k$ solutions exist; a (conceptual) nondeterministic Turing machine requires polynomial time since it checks all inputs in parallel; and a quantum computer applying Grover's algorithm will require time $O(\sqrt{2^n/k})$, a quadratic speedup over classical approaches.  To illustrate the advantage of this quadratic speedup, consider a problem with an $n=32$-bit input where a single solution exists ($k=1$), and assume that classically computing $f()$ on an input takes 10 ms.  Checking half the inputs (the expected number to find the solution) would require $2^{31} / 100$ seconds, which is just under 250 days.  In contrast, even if the corresponding overhead of computing $f()$ in the context of Grover's algorithm (high-level details below) took 1 second each (a relative slowdown of 100 per instance), a solution could be found in slightly over 18 hours.  Were $n=48$, the respective times would be just under 450,000 years for classical vs. 194 days for quantum.

\begin{figure}[!htb]
\includegraphics[width=0.45\textwidth]{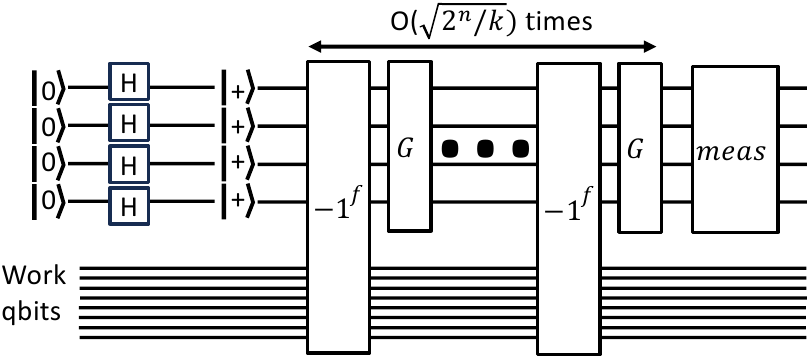}
\caption{Grover overview}
\label{fig:grover}
\end{figure}

The general process of Grover's algorithm is depicted in Figure \ref{fig:grover}.  Grover's algorithm starts much like a non-determin\-ist\-ic Turing machine, generating (in superposition) the set of all possible inputs (the time to do this is linear in the size of the input): the $n$ qubits, initially set in a ground state (i.e., to 0 or more formally, $\ket{0}$), are each operated upon by a Hadamard gate that shifts each qubit to a state known as $\ket{+}$ which, if measured, returns 0 or 1 with equal probability of $0.5$.  This superimposed state is simultaneously run through the verifier circuit, $f$, which, in superposition, computes $f$ on all $2^n$ inputs.  The output of $f$ is applied (in superposition) as the exponent to $-1$ (to produce 1 when $f=0$ and $-1$ when $f=1$), thereby generating (in superposition) the result of each input (success or failure of the desired result).  A final step is to choose one of the $k$ inputs that was successful, but unfortunately this part is not easy, and extracting an appropriate solution is what adds the $O(\sqrt{2^n/k})$ additional complexity.  This extraction process involves repeated application of a sequence of gate operations often termed a Grover iterate $G$, which combines $f$ with a {\em Diffuser} circuit, as shown in Figure \ref{fig:grover}, after which, the $n$ qubits that represented the input are measured.  With probability close to 1, these bits will collapse to an $n$-bit input that matches an input for which $f$ evaluated to 1.  In effect, this $O(\sqrt{2^n/k})$ extraction process finds a desired input \cite{Nielsen_Chuang_2010}.  In this sense, it differs from a nondeterministic Turing machine in two ways: first, it has an additional $O(\sqrt{2^n/k})$ overhead to find a solution, and it only returns a single solution, effectively selected at random, as opposed to conceivably returning all possible solutions by the end of the (nondeterministic) run.

\section{Mapping NWV to Quantum Unstructured Search}
\label{sec:MappingNVtoQuantum}

\begin{figure}[!htb]
\includegraphics[width=0.45\textwidth]{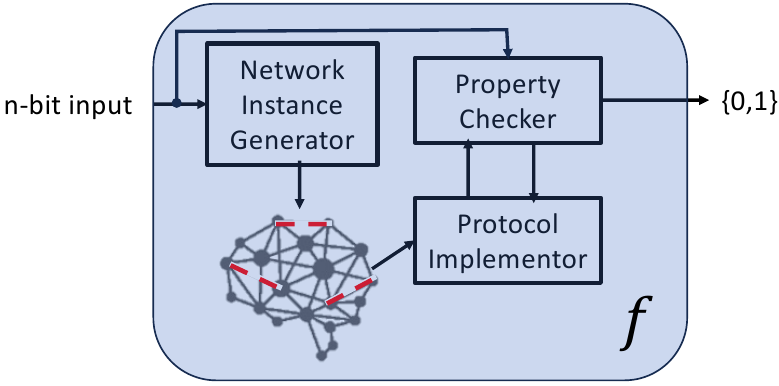}
\caption{Components of the oracle function $f$}
\label{fig:f-construction}
\end{figure}

In this section, we describe how to map Grover's algorithm to a generalized instance of a NWV problem.  In short, we must define an oracle function $f$ as a {\em verifier} that takes an $n$-bit input to specify a specific instance of the search space and emulates network operation on that instance to validate whether the desired (or undesired) property holds for that instance.  The basic functionality that $f$ must perform is depicted in Figure \ref{fig:f-construction}.  Given an $n$-bit input instance, $f$
\begin{myitemize}
    \item generates a network whose configuration can be influenced by the $n$-bit input, which we call a {\em network instance}. 
    \item then runs the underlying routing protocol upon the network instance.
    \item in parallel to the execution of the protocol, a {\em property checker} checks whether the property holds or is violated for the current network instance.
    \end{myitemize}
     By encoding $f$ as a circuit that can be executed in a quantum computer, Grover's algorithm can be applied to simultaneously evaluate all possible $2^n$ instances.

In the following two subsections, we separately describe application of the general mapping approach to NWV problems to the Data Plane and Control Plane respectively, each followed by a very preliminary, proof-of-concept mapping of the problem which we analyzed on existing NISQ quantum hardware.

\subsection{Data Plane Example}

Consider a data plane setting similar to that proposed in \cite{kazemian2012header}, again viewing the underlying network as a graph, next hops are determined by the header in a packet, with the $n$-bit packet header possibly modified (in a deterministic manner) as the packet routes through the network.  We wish to determine if, from a specific source node $S$, whether there are any packets that will traverse more than 100 hops.  This property can be hard to detect since cycles are permissible that can be exited from due to the changing nature of the packet header as the packet traverses the network.  We design $f$ as follows:

\begin{myitemize}

\item The {\em Network Instance} contains a hard-coded version of $G$, as well as a hard-coded set of forwarding rules and packet header modification routines for each node in the network.  The $n$ bit input indicates the packet header, and is the only variable component to the network instance.

\item This graph is passed into the {\em Protocol Implementer} which emulates the hard-coded forwarding rules upon the provided $n$-bit packet header for 100 hops. 

\item The {\em Property Checker} monitors the path taken by the packet as it traverses through the network, counting the number of hops as it proceeds along for at least 100 hops.  If the packet reaches its destination prior to the 100th hop, a 0 is immediately returned.  Otherwise, upon reaching the 100th hop, a 1 is returned.

\end{myitemize}

\subsection{Control Plane Example}
Consider a control plane setting similar to that proposed in \cite{netdice} in which a given underlying network can be viewed as a graph $G=\{V,E\}$ with $n = |E|$, the number of links in the network.  A routing protocol emulating BGP/IGP determines routes through this graph.  Selecting specific nodes $C,D,E \in V$ on this network, we wish to evaluate whether there exist combinations of up to 10 link failures for which $C$'s path to $D$ fails to proceed through node $E$.  We design $f$ as follows:

\begin{myitemize}
    \item The {\em Network Instance Generator} contains a hard-coded version of $G$, and uses the $n$-bit input to indicate (1 bit per link) the subset of the $n$ links that fail, removing these links from the instance of the graph.

    \item This revised graph is passed to the {\em Protocol Implementer} which proceeds to emulate a hard-coded implementation of BGP/IGP on the underlying network.

    \item The {\em Property Checker} then explores the route from $C$ to $D$, returning 0 as soon as vertex $E$ is reached (i.e., before $D$).  Also, if the $n$-bit input itself initially has more than 10 bits encoded to 1 (i.e., more than 10 link failures), 0 is also returned.  Otherwise, if $D$ is reached without going through $E$, then 1 is returned.
\end{myitemize}

In addition to implementing a "cutoff" that bounds the maximal number of failing links, a variant of Grover known as {\em Amplified Grover} can be utilized to increase the efficiency of finding solutions that center around a specific number of link failures.  While the traditional implementation of Grover applies the Hadamard gate on each qubit to put its measurement outcome split in half between 0 and 1, an enhanced {\em Amplified Grover} can instead initialize the state of each qubit into any arbitrary probability $p$ of evaluating to 0 (link failure).   Doing so changes the relative likelihoods of Grover selecting a given outcome, maximizing those centered around the distribution where a fraction $p$ of the links fail.  Hence, in a large network where we are interested in only considering the cases where small numbers of links are expected to fail, we can utilize values of $p$ that are far smaller than 0.5.

\section{(Very) Preliminary Implementation}

\begin{figure}[!htb]
\begin{subfigure}{0.4\columnwidth}
\includegraphics[width=1.0\columnwidth]{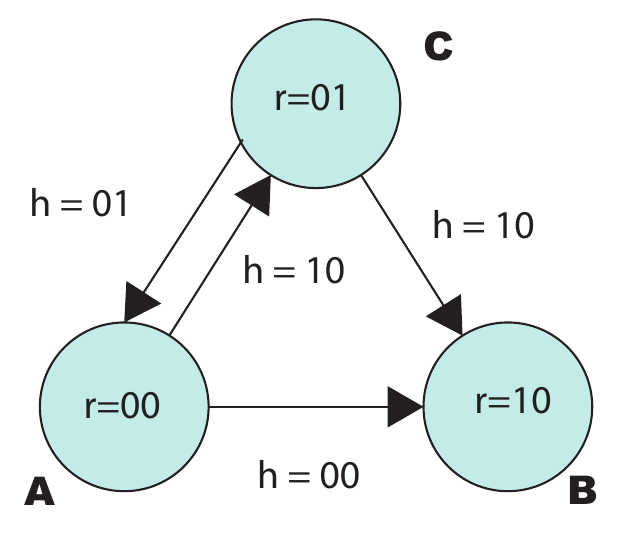}
\caption{Data Plane}
\label{fig:toyHSA-DataPlane}
\end{subfigure}
\hfill
\begin{subfigure}{0.4\columnwidth}
\includegraphics[width=1.0\columnwidth]{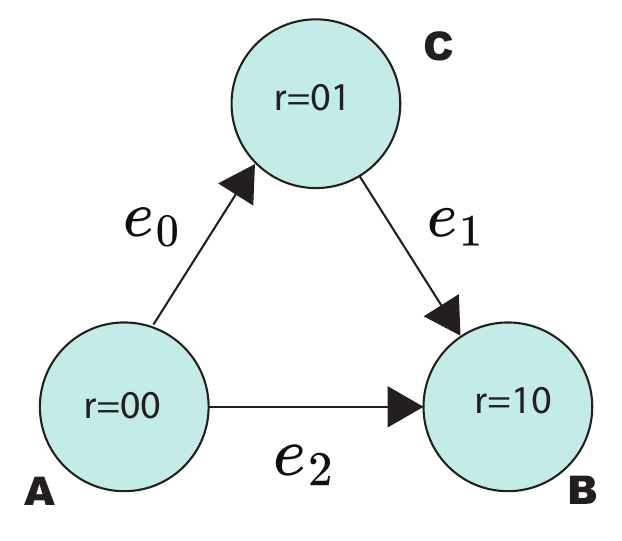}
\caption{Control Plane}
\label{fig:toyHSA-ControlPlane}
\end{subfigure}
\caption{Example Networks for Experiments}
\label{fig:toyHSA}
\end{figure}

For purposes of proof-of-concept and to evaluate initial efficacy of using quantum computation, we consider a very simple network that can be easily analyzed on today's QC platforms.  Note that our examples are merely meant to be illustrative of how to perform NWV on top of QC: the hardware available to us is nowhere near at scale such that we observe quantum advantage.  In fact, our networks are so simplistic that validation can simply be done by hand.  Figure \ref{fig:toyHSA} depict simple 3-node networks (with nodes A,B,C) with 2-bit router IDs $r$ assigned.  Figure \ref{fig:toyHSA-DataPlane} depicts a network where routers forward packets with 2-bit headers in the directions indicated by arrows (dataplane example), while in Figure \ref{fig:toyHSA-ControlPlane}, we annotate 3 edges $e_0, e_1,$ and $e_2$ which may fail, and evaluate whether flows emanating from router node $A$ can reach node $C$.

\subsection{Data Plane Proof-of-concept}
\label{sec:hsademo}
We demonstrate an example of mapping NWV to Quantum Search for the Data Plane setting.  Figure \ref{fig:toyHSA} shows a toy network with 3 routers, A, B and C.  We assume packets have two-bit headers.  The arrows represent the network routing behavior, which can be described similarly to the transfer functions of Header Space Analysis; arrow labels detail the next-hop behavior for headers.  Headers with no corresponding arrows remain at their router without being forwarded.
\begin{figure}
    \centering
    \includegraphics[height=0.4\textheight]{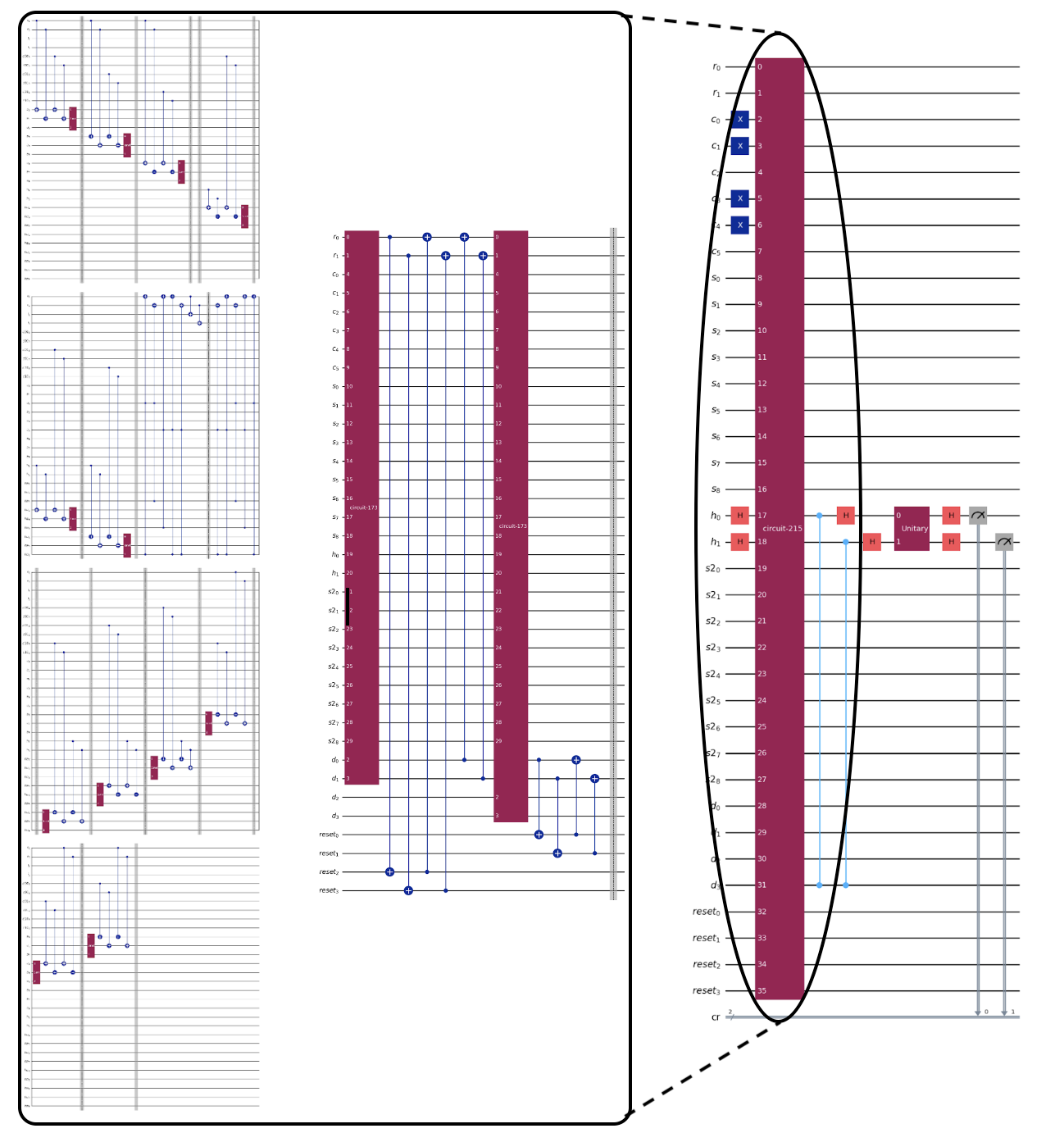}
    \caption{Left: Detail of Quantum Circuit to implement transfer function logic.  Middle: Oracle circuit $f$ of Grover's Algorithm.  Right: Full circuit for Quantum Search Algorithm, run for one Grover iterate.}
    \label{fig:hsa_qc}
\end{figure}

We consider checking the property of 2-bit headers must route from A to C within 2 hops (i.e., avoiding loops).  To construct the quantum circuit, the network instance generator and protocol implementer, collectively referred to as the oracle $f$, are simultaneously integrated into one quantum circuit block\TROnly{, as shown in the left and middle of Figure \ref{fig:hsa_qc}.}{ (please see \cite{OurTR} for more details on the quantum circuit).}  We associate two bits to each router represent the router location.  The "input" that $f$ takes is four bits long: two for the header, which will be put into superposition for Grover's Algorithm, and two to represent the “current router location,” which is reset in between iterations. 

Our quantum circuit employs standard combinatorial logic to implement the circuit and its routing behavior. First, the header bits of the input are fed through a "bit checker"; conditional on their value,  the input “current router location” bits are modified, in line with the network forwarding logic.  \TROnly{This part of the circuit is shown in detail in the left of Fig. \ref{fig:hsa_qc}.}{}  

The circuit consisting of bit-checking and altering router location represents one "hop" of the routing process; to simulate a second hop, this circuit is applied once more, with all inputs except the "current router location" reset to their original value.  \TROnly{This is shown in the middle of Figure \ref{fig:hsa_qc}.}{}

\textbf{Property Checker:} After applying the above circuit to simulate two "hops" of routing logic, the property checker is another logical circuit that simply checks the value of the "current router location" bits to see if they are set to "$10$" (representing router C).  \TROnly{Controlled on the logical outcome being true, a pauli-Z gate is applied to the header bits; this is the "phase-marking" part of Grover's algorithm. This is shown in the right of Figure \ref{fig:hsa_qc}.}{}

The combined Network Instance Generator and Protocol Implementer represent the "oracle" of Grover's algorithm.  For the full algorithm, the header bits of the input are initially put into equal superposition, and fed into the Grover oracle.  For this problem, a total of one Grover iterates (oracle + diffusion) are sufficient to extract the solution headers ("$00$" and "$10$") with high probability.  \TROnly{The full circuit implementing this Grover iterate is shown on the right of Figure \ref{fig:hsa_qc}.}{}

\begin{figure}[!htb]
    \centering
    \begin{subfigure}{0.45\columnwidth}
    \includegraphics[width=1.0\textwidth]{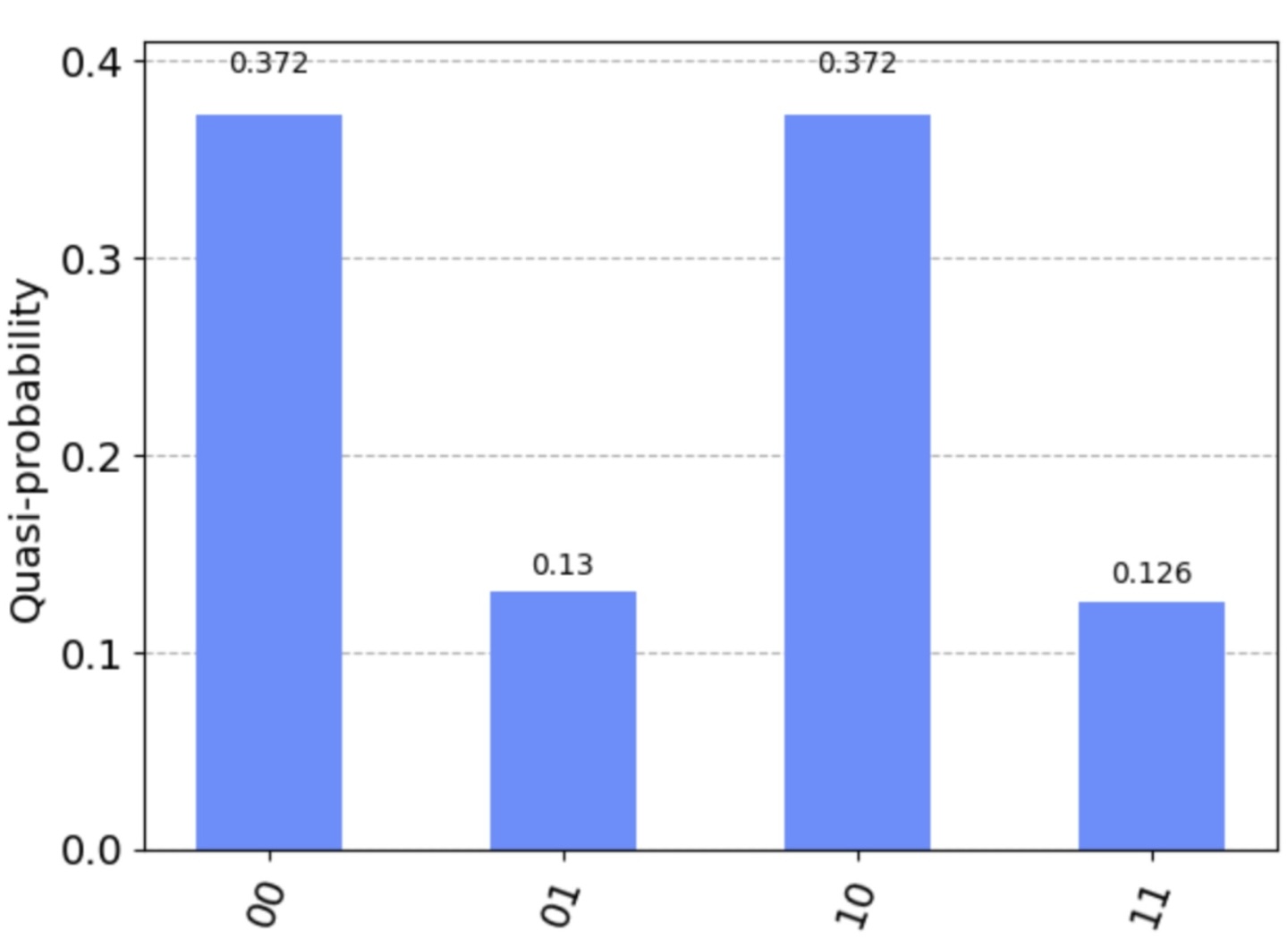}
    \caption{simulation}
    \label{fig:hsa_hist_sim}
    \end{subfigure}
    \hfill
    \begin{subfigure}{0.45\columnwidth}
    \includegraphics[width=1.0\textwidth]{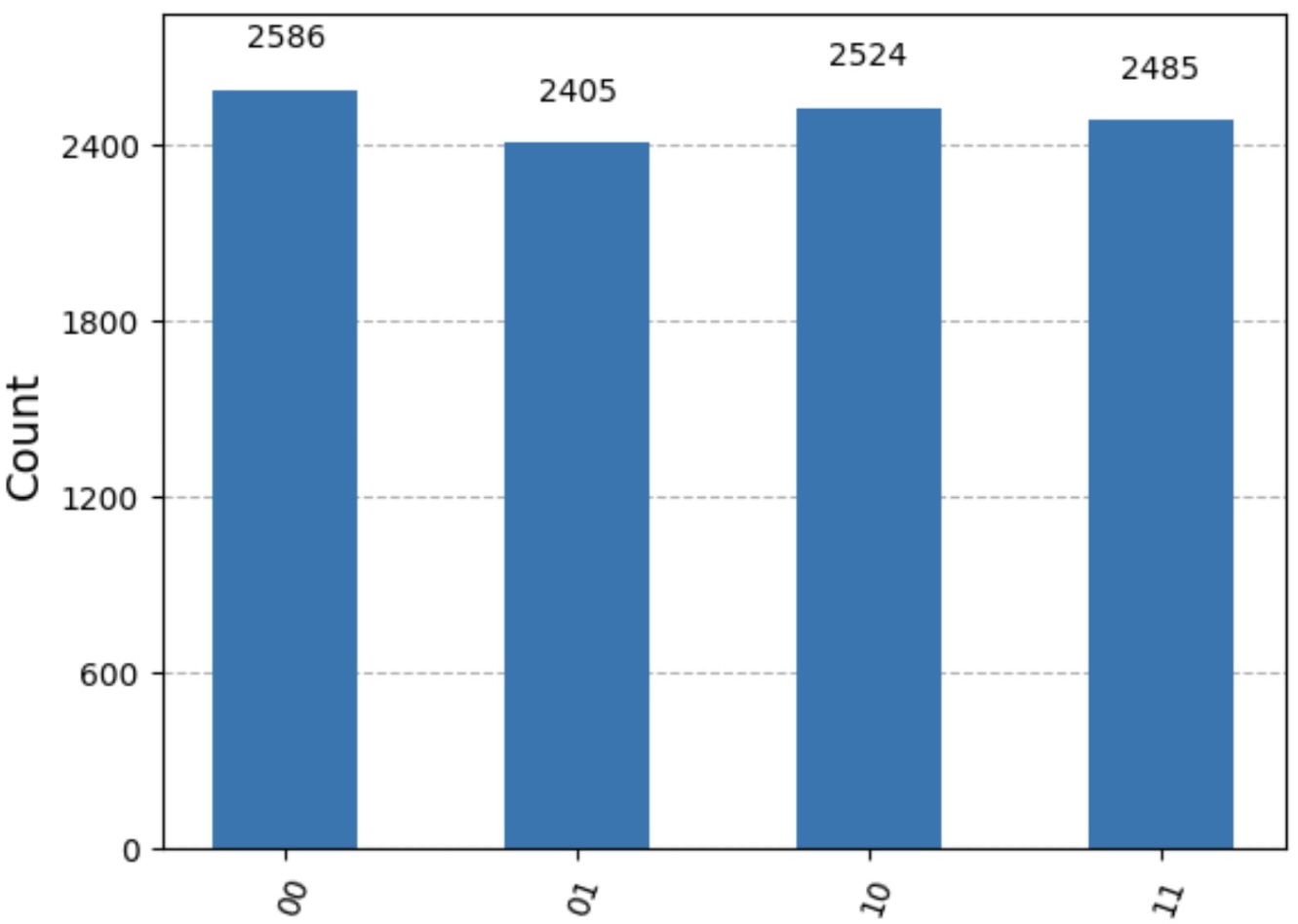}
    \caption{experimentation}
    \label{fig:hsa_hist_exper}
    \end{subfigure}
       \caption{10,000 shots of Grover's Algorithm} \label{fig:hsa_hist}
\end{figure}

We implemented this circuit in IBM Qiskit, and ran it on both IBM's  AerSimulator and on IBM's Osaka, an actual 127-qubit Quantum Computer.    Figure \ref{fig:hsa_hist} shows the end results for both.  When simulated (Figure \ref{fig:hsa_hist_sim}), Grover's algorithm returns a correct header roughly $75\%$ of the time.  Ironically, Grover's algorithm increases in accuracy as the size of the input grows.\footnote{Each iteration can be viewed as a rotation, where the size of the rotation angle is inversely proportional to the ratio of solution inputs to the space of all inputs.  Hence, when the space of all inputs is small, one is forced to either significantly over- or under-rotate.}  We note, however that for small inputs, with enough repeated runs of the entire circuit, we can ensure we obtain a correct answer with high probability.  On the real quantum machine (Figure \ref{fig:hsa_hist_exper}), this correct proportion is reduced to $51\%$, additional demonstration that quantum computation is not yet ready for conventional use.

\subsection{Control Plane Proof-of-Concept}

Using the topology shown in Figure \ref{fig:toyHSA-ControlPlane}, we  consider all 8 possible combinations of link outages.

\begin{figure}[!htb]
    \centering
    \includegraphics[width= 0.8\columnwidth]{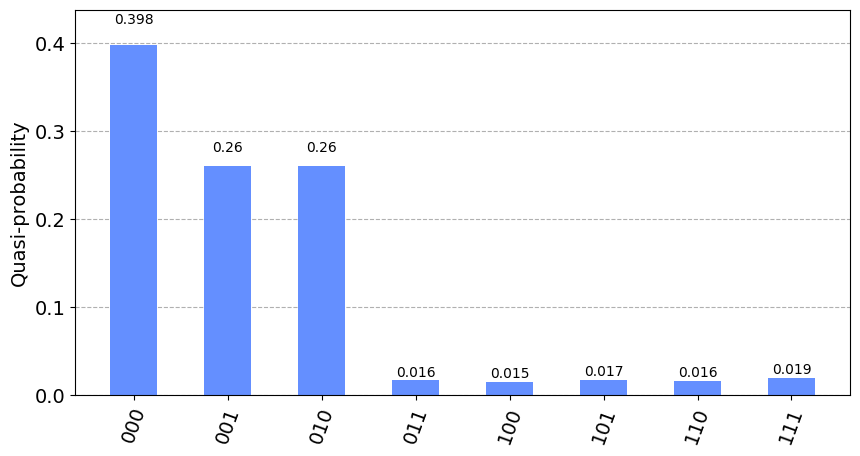}
    \caption{Results of Control Plane Analysis}
    \label{fig:netdice_hist}
\end{figure}

\begin{figure*}
    \centering
    \includegraphics[scale=0.3]{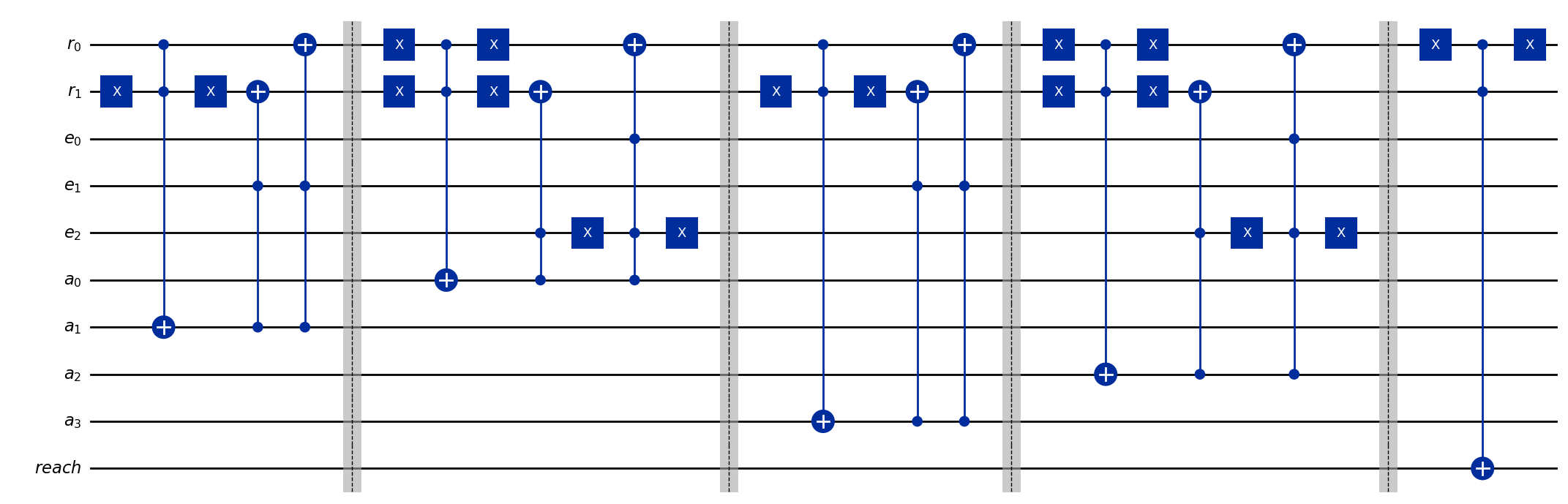}
    \caption{Link Outage Oracle Circuit}
    \label{fig:netdice_qc}
\end{figure*}

\TROnly{Figure~\ref{fig:netdice_qc} depicts the quantum circuit that}{Our quantum circuit} implements $f$ to return 1 when the destination at $C$ is unreachable from the source at $A$.   Figure \ref{fig:netdice_hist} shows histogram results of outputs when applying Grover, where the label $x_2 x_1 x_0$ indicates the status of respective links $e_2, e_1,$ and $e_0$ where 1 indicates the link is operational (0 indicates failure).  The most frequently returned results represent all failure scenarios.


\newcommand{\subfigsize}{0.33\textwidth}

\section{Quantum Circuit Complexity Analysis}
\begin{figure*}
    \centering
    \begin{subfigure}{\subfigsize}
    \includegraphics[width= 1.0\columnwidth]{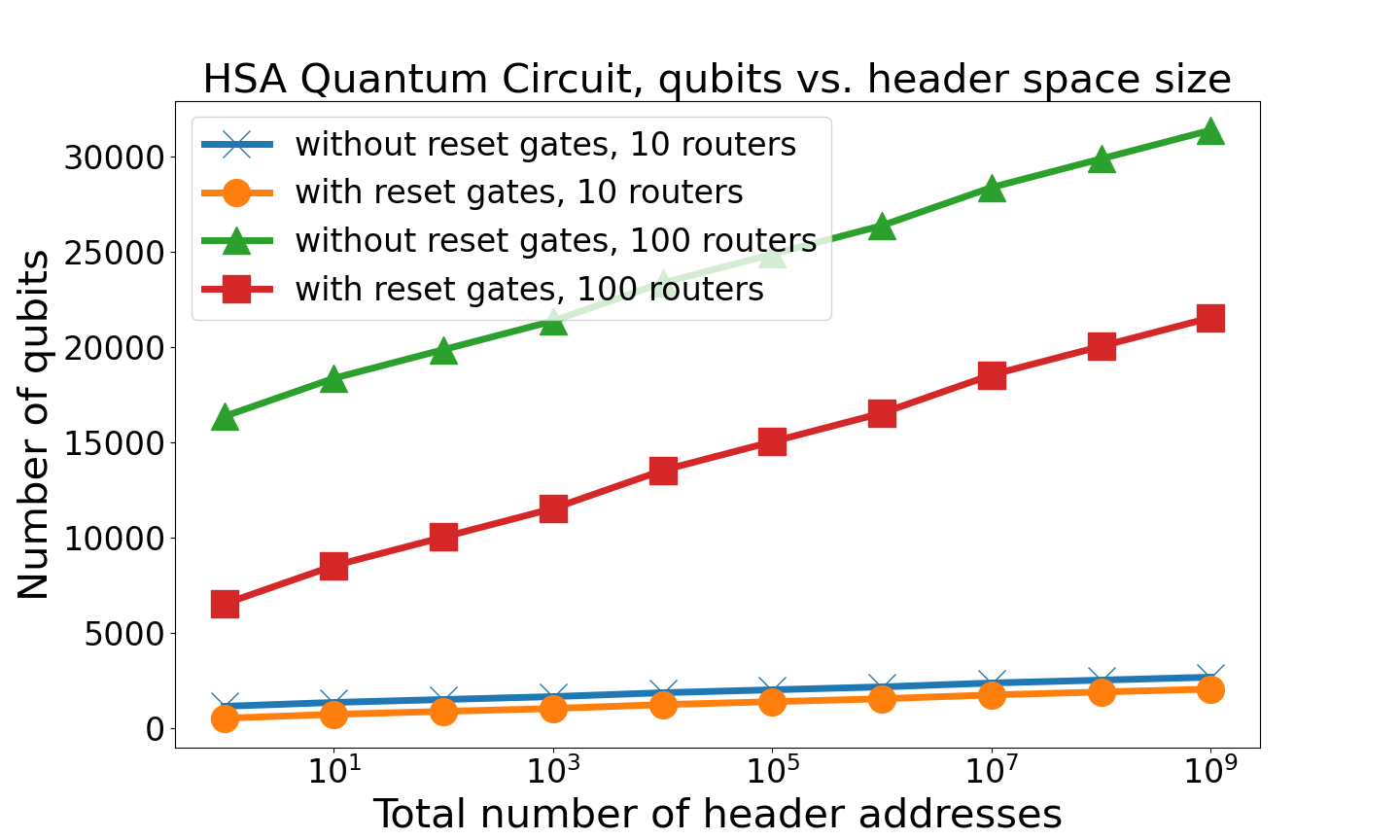}
    \caption{Data Plane varying size of input $n$}
    \label{fig:hsa_qubits_vs_L}
\end{subfigure} \hfill
\begin{subfigure}{\subfigsize}
    \centering
    \includegraphics[width= 1.0\columnwidth]{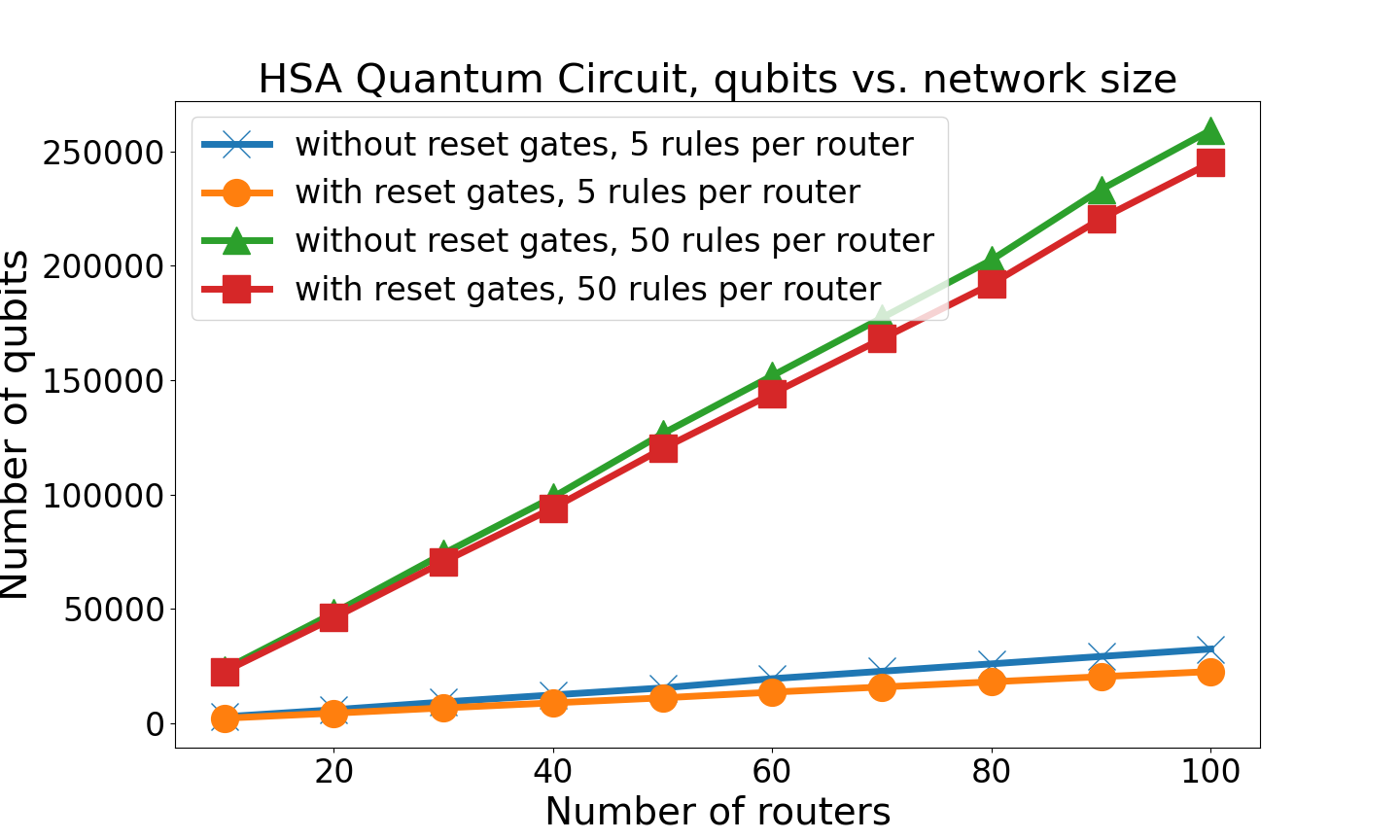}
    \caption{Data Plane varying number of routers}
    \label{fig:hsa_qubits_vs_R}
\end{subfigure} \hfill
\begin{subfigure}{\subfigsize}
    \centering
    \includegraphics[width= 1.0\columnwidth]{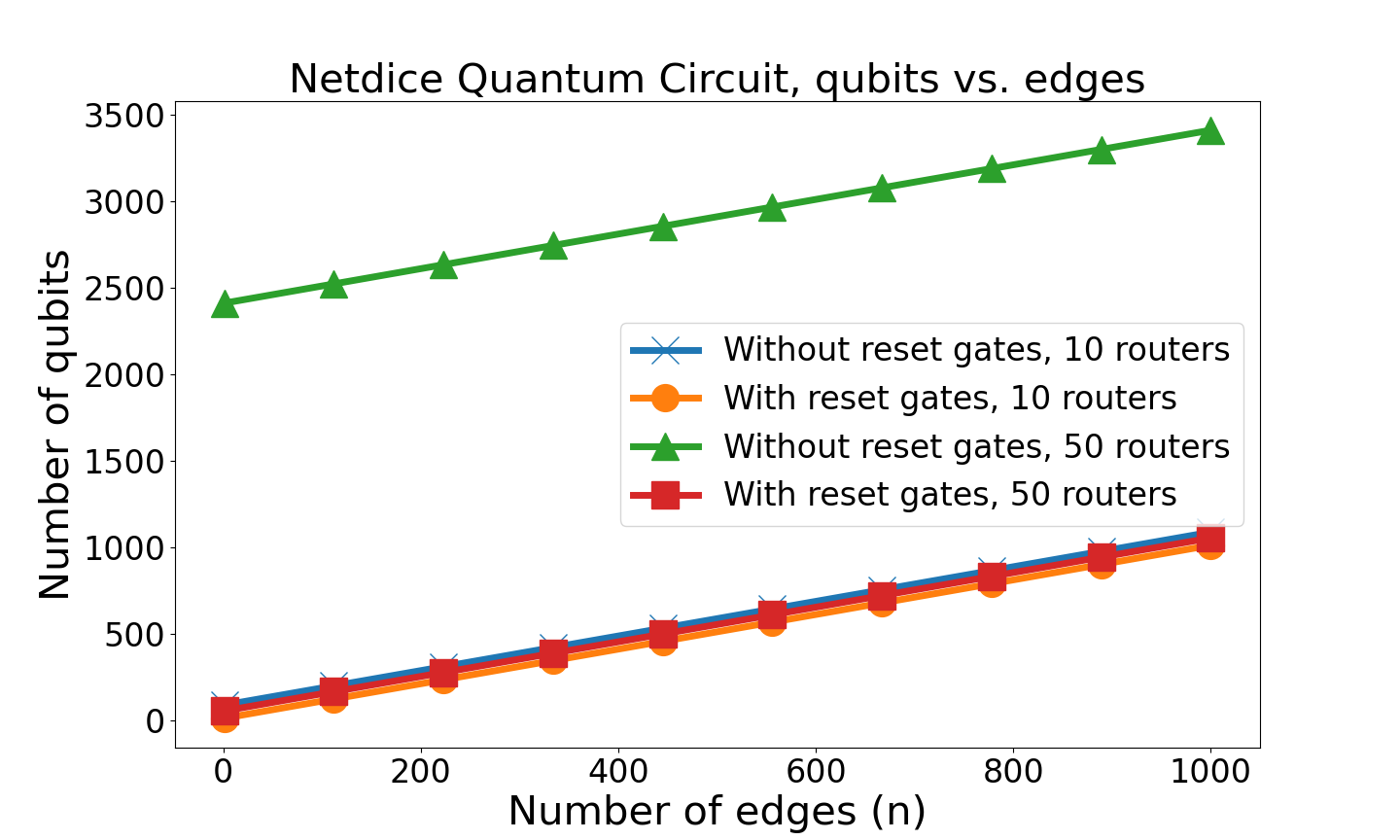}
    \caption{Control Plane varying size of input $n$}
    \label{fig:nd_qubits_v_n}
    \end{subfigure}
    \caption{Required number of qubits for implementation}
\end{figure*}

The time required to perform NWV is dependent on many factors for (error correction codes and their corresponding rates, as well as the general speed of the circuits and how well these circuits can be optimized).  It is difficult at this time to presuppose the time computation will take.  We can, however, speculate on the number of qubits that a computation will require as a function of the size of the problem being considreed.  In this section, we make an attempt to quantify how the number of required quantum resources of our circuits (in terms of number of qubits) scales with the size of the problem.  

\subsection{Data Plane Verification}

For Data Plane Verification with Header Space Analysis, we consider two ways to quantify the size of the problem: the  number of possible header addresses and the size of the network (in terms of the number of rules and routers). 

For simplicity, we consider a network that consists only of if/then rules matching a header wildcard expression and forwarding to a particular port number.  Define $n$ to be the total number of headers, $R$ the total number of routers, $\ell$ the number of unique wildcard expressions that appear in the rules of the network, $P$ the number of unique port numbers that appear in the rules, $k$ the max number of hops, and $G$ the optimal number of Grover iterates.  A simple calculation shows that, in terms of these variables, our Quantum Circuit requires $(1+\ell)\lceil \log(n) \rceil + (P + k + G(2k - 1))\lceil \log(P) \rceil + 2 \max(\ell,P) + P + \ell$ qubits.  If we allow the use of mid-circuit reset gates (instead of using extra ancilla qubits), we require only $(1+\ell)\lceil \log(n) \rceil + (1+P)\lceil \log(P) \rceil + 2 \max(\ell,P) + P + \ell$.  

Figure \ref{fig:hsa_qubits_vs_L} shows how the number of required qubits varies with $n$ (varied logarithmically on the $x$-axis), for network sizes of $10$ and $100$ routers.  For ease of analysis, we assume every router has the same number of rules $r$, and $\ell = P = R \cdot r$. We also set $k = R$ and $G=5$.  Unsurprisingly, the number of qubits required scales linearly with the log of the number of headers (i.e., linearly with the number of input bits).

Figure \ref{fig:hsa_qubits_vs_R} shows how the required number of qubits varies with $R$, with the number of rules per router $r$ set to 5 and 50. Again, we set $\ell = P = R \cdot r, k = R$ and $G = 5$.  We see the scaling is linear in $R$, which is to be expected as we have set $R$ to be directly proportional to both $\ell$ and $P$.  The number of qubits needed for computation scales linearly in the size of the network.

\subsection{Control Plane Verification}
For Control Plane Verification with NetDice, we quantify the size of the problem in terms of the input space $n$, the total number of edges in the network. We define $R$ to be the total number of routers, $D$ to be the diameter of the network, and $G$ the optimal number of Grover iterates. In these terms, this quantum circuit requires $\log(r) + e ((r-1)*d) + g$ total qubits. Allowing mid-circuit reset gates reduces this number to $\log(r) + e + r$ qubits. 

Figure \ref{fig:nd_qubits_v_n} shows how the number of required qubits varies with $n$, for network sizes of $10$ and $50$ routers. We assume $d = r - 1$ and $g = r$.  Again, we see a linear scaling in the size of the input.
\section{Related Work}

\textbf{Quantum Computing for Program Verification:}
To our knowledge, our work is the first to propose the use of quantum computing for network verification.  Recent work explores quantum computing for program verification~\cite{issel2024towards}.  Due to aforementioned complexities in verification of general programs, the program cannot be analyzed directly as in our case, but must be converted to a corresponding SAT instance.  

\textbf{Network Verification:} There are a wide range of classical approaches and techniques to verifying network properties have been developed  \cite{8453007}.  In general, approaches to NWV can be divided into two classes.   {\em Control Plane verification}  (\cite{fastplane,anteater, tiramisu,minesweeper, netdice}) analyses the network at the level of protocol and device configurations, derives the subsequent routing behavior, and then checks the relevant properties. Holistic approaches intend to holistically represent *all* possible data planes induced by a given control plane configuration via symbolic representation but at the expense of computational tractability \cite{minesweeper, arashloo2023formal, prabhu2020plankton, sherwood2024netcastle}. On the other hand, "instance-based" approaches emulate the convergence process from a given control plane to a fixed data plane, but they face the challenge of accurately modeling every network component and risk missing critical corner cases \cite{brown2023lessons, batfish_old, tiramisu}. {\em Data Plane verification} (\cite{kazemian2012header,anteater,al2010flowchecker,zhang2013sat}) assumes the network routes to have already been established, and checks relevant properties by analyzing the network forwarding tables. These methods generally aim to group packet headers with identical routing behavior for optimizing the search of instances that might violate the property of interest. However, the number of groups can increase exponentially with the number of network components, packet headers, and middleboxes. To maintain efficient verification, these techniques make assumptions about a given network's underlying structure. When violated, the claims to efficiency may no longer hold.   


 \section{Discussion / Future Plans}

This paper's contribution is intended to introduce to the networking community an application of quantum computing to a computationally challenging problem in networks: that of network verification.  Our preliminary work is meant to demonstrate that there will be certain challenges in network verification which remains open and impractical to solve using classical computing.  Instead, by phrasing the NWV problem as a functional verifier, quantum's ability to speed up unstructured search may itself provide sufficient benefit to make these challenges verifiable in reasonable time.

Our initial foray indicates that there is still a significant amount of work to be done.  First, building the verifier $f$ for large networks and sophisticated properties remains a challenging endeavor: for now it must be designed at the circuit level.  Second, our experience with Grover's algorithm thusfar shows that its performance in practice is noisy (too often returning instances where the property was not violated).  This overhead must also be taken into account, or somehow approved upon.  Last, the speedup gained by utilizing unstructured search, while significant, does not take full advantage of the potential power of Quantum Computing.  It remains to try and understand how existing structure in the underlying problem (in this case, NWV) can be exploited to utilize approaches that can potentially have exponential speedup when performed upon a Quantum system.
 
\section*{Acknowledgements}
We wish to thank our colleague Henry Yuen and the anonymous HotNets reviewers for their insightful feedback on earlier drafts. 
 This material is based upon work supported by the National Science Foundation under Grant No. CNS-2148275.  Any opinions, findings, and conclusions or recommendations expressed in this material are those of the author(s) and do not necessarily reflect the views of the National Science Foundation.

\bibliographystyle{ACM-Reference-Format} 
\bibliography{QNV-hotnets24}

\end{document}